\documentclass[letterpaper]{article} 
\usepackage{aaai25}  
\usepackage{times}  
\usepackage{helvet}  
\usepackage{courier}  
\usepackage[hyphens]{url}  
\usepackage{graphicx} 
\usepackage{natbib}  
\usepackage{caption} 
\usepackage{algorithm}
\usepackage{algorithmic}

\usepackage{newfloat}
\usepackage{listings}
\DeclareCaptionStyle{ruled}{labelfont=normalfont,labelsep=colon,strut=off} 
\floatstyle{ruled}
\newfloat{listing}{tb}{lst}{}
\floatname{listing}{Listing}
\usepackage{booktabs}
\usepackage{color}
\usepackage{adjustbox}
\usepackage{array}
\usepackage{xcolor,colortbl}
\usepackage{appendix}
\usepackage{algorithm,algorithmic}
\usepackage{subfig}
\usepackage[countmax]{subfloat}

\usepackage{multirow}
\usepackage{amsmath}
\usepackage{amssymb} 
\newcommand{\argmin}{\arg\!\min}
\newcommand{\Amat}[0]{{{\textbf A}}}

\newcommand{\Fmat}[0]{{{\textbf F}}}

\newcommand{\Xmat}{{\textbf X}}
\newcommand{\Ymat}[0]{{{\textbf Y}}}
\newcommand{\Zmat}{{\textbf Z}}

\newcommand{\bv}{\boldsymbol{b}}

\newcommand{\dv}{\boldsymbol{d}}

\newcommand{\rv}[0]{{\boldsymbol{r}}}

\newcommand{\xv}{\boldsymbol{x}}
\newcommand{\yv}{\boldsymbol{y}}

\newcommand{\etav}[0]{{\boldsymbol{\eta}} }
\newcommand{\thetav}{\boldsymbol{\theta}}

\newcommand{\muv}{\boldsymbol{\mu}}

\newcommand{\phiv}{\boldsymbol{\phi}}

\usepackage{xspace}
\def\xnet{DG-NLOS\xspace}

\title{Dual-branch Graph Feature Learning for NLOS Imaging}
\author{
Xiongfei Su\textsuperscript{\rm 1,2,3}\equalcontrib, Tianyi Zhu\textsuperscript{\rm 2}\equalcontrib, Lina Liu\textsuperscript{\rm 2}, Zheng Chen\textsuperscript{\rm 4}, Yulun Zhang\textsuperscript{\rm 4}, \\
Siyuan Li\textsuperscript{\rm 1,3}, 
Juntian Ye\textsuperscript{\rm 5}, Feihu Xu\textsuperscript{\rm 5}, Xin Yuan\textsuperscript{\rm 3}\thanks{Corresponding author.}
}
\affiliations{
    \textsuperscript{\rm 1}Zhejiang University, Hangzhou, China, e-mail: xsuac@zju.edu.cn \\
    \textsuperscript{\rm 2}China Mobile Research Institute, Beijing, China \\
    \textsuperscript{\rm 3}Westlake University, Hangzhou, China\\
    \textsuperscript{\rm 4}Shanghai Jiao Tong University, Shanghai, China \\
\textsuperscript{\rm 5}University of Science and Technology of China, Anhui, China

}

\usepackage{bibentry}

\begin{document}

\maketitle

\begin{abstract}
The domain of non-line-of-sight (NLOS) imaging is advancing rapidly, offering the capability to reveal occluded scenes that are not directly visible. However, contemporary NLOS systems face several significant challenges: (1) The computational and storage requirements are profound due to the inherent three-dimensional grid data structure, which restricts practical application. (2) The simultaneous reconstruction of albedo and depth information requires a delicate balance using hyperparameters in the loss function, rendering the concurrent reconstruction of texture and depth information difficult. This paper introduces the innovative methodology, \xnet, which integrates an albedo-focused reconstruction branch dedicated to albedo information recovery and a depth-focused reconstruction branch that extracts geometrical structure, to overcome these obstacles. The dual-branch framework segregates content delivery to the respective reconstructions, thereby enhancing the quality of the retrieved data. To our knowledge, we are the first to employ the GNN as a fundamental component to transform dense NLOS grid data into sparse structural features for efficient reconstruction. Comprehensive experiments demonstrate that our method attains the highest level of performance among existing methods across synthetic and real data. https://github.com/Nicholassu/DG-NLOS.

\end{abstract}

%

\section{Introduction}
Non-line-of-sight (NLOS) imaging transcends the limitations of rectilinear propagation by examining the diffuse reflections from a relay surface to image occluded objects~\cite{faccio2020non,maeda2019recent}. This technique has extensive applications across various domains, including medical imaging, autonomous driving, and robotic vision.
The swift progression in photon-sensitive sensors and sophisticated imaging algorithms has engendered solutions that span a comprehensive range of ray propagation paradigms, such as occluder-based recovery~\cite{xu2018revealing,saunders2019computational}, Time-of-Flight (ToF) methodologies~\cite{velten2012recovering,heide2014diffuse}, 
wavefront shaping~\cite{cao2022high}, and alternative non-optical techniques~\cite{lindell2019acoustic,scheiner2020seeing}. Several methods exist for reconstructing hidden scenes, including Fermat flow~\cite{xin2019theory}, temporal focusing~\cite{pediredla2019snlos}, optimization-based reconstruction~\cite{heide2019non,tsai2019beyond,ye2021compressed}, deconvolution~\cite{otoole_confocal_2018,young2020non},
wave-based techniques~\cite{liu2019non, lindell_wave-based_2019}, and deep learning~\cite{grau2020deep,nam2021low,pei2021dynamic,liu2022hiddenpose, ye2024plug}.
\begin{figure}
    \centering
    \includegraphics[width=\linewidth]{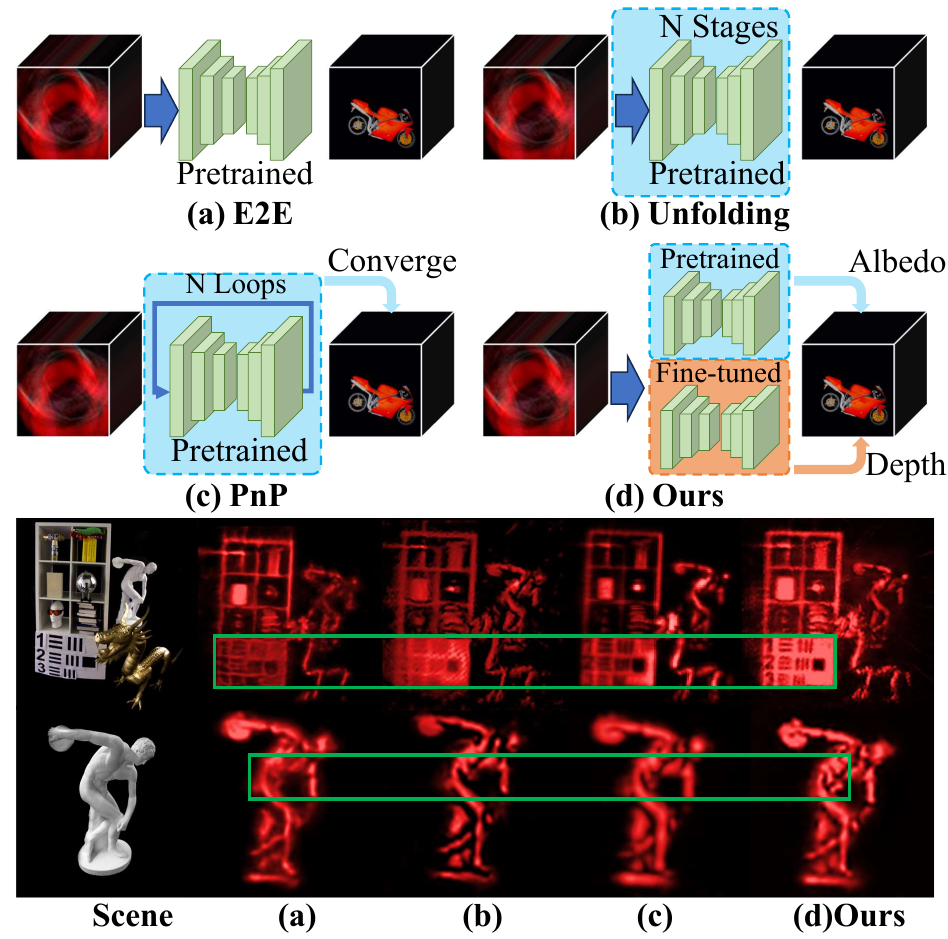}
    \caption{Various types of Deep NLOS reconstruction. \textcolor{blue}{Blue} arrows denote the transform with physical models. The two real scene reconstructions are compared, where \textcolor{green}{green} rectangles highlight the superior performance of \xnet. 
    }
    \label{fig:types}
\end{figure}

NLOS reconstructions commonly face the following challenges: \textbf{($\textit{\textbf{i}}$) Low signal-to-noise ratio (SNR):} as only a small fraction of scattered photons are captured~\cite{zhu2023compressive, li2024deep}. \textbf{(\textit{ii}) Ill-posed solution:} Signals travel through multiple paths to reach the receiver, causing time and phase differences between the paths~\cite{wt_dense_2021}, resulting in multipath interference. \textbf{(\textit{iii}) Complex signal detection and estimation:} NLOS environments require complex signal processing algorithms~\cite{yu2023enhancing, liu2023non}, channel estimation methods, and modulation/demodulation techniques to handle the multipath effects, angle spread, and time-varying characteristics.

Three deep learning-based frameworks are proposed to solve these problems, shown in Fig.~\ref{fig:types}: \textbf{(a) End-to-end (E2E):} \cite{chen2020learned} introduces a feature embedding method for NLOS tasks, including reconstruction, imaging, classification, and object detection. \cite{grau2020deep} generates training images using noise and rendering models for transient NLOS imaging with Time-of-Flight (ToF) technology, modifying U-Net with a 3D tensor to convert transient measurements into depth maps.~\cite{optica_deep_inverse} use the spectral estimation theory for NLOS correlography, employing a deep neural network to handle noisy phase retrieval and detect hidden objects through indirect reflections. \cite{li_nlost_nodate} introduces Transformer to extract global and local relationship. \
\textbf{(b) Deep Unfolding} and \textbf{(c) Plug-and-Play (PnP)} explore hybrid approaches combining the benefits of both optimization and deep learning paradigms. Deep Unfolding~\cite{su2023multi} trains a sequence of interconnected small sub-networks, emulating the iterative process of traditional optimization. This method ``unfolds" optimization iterations with end-to-end training and offers interpretability by mapping stages to iterations. PnP algorithms~\cite{ye2024plug} use pre-trained deep denoising networks as priors in iterative algorithms, requiring no additional training for new applications. They are adaptable to various systems and ensure algorithm stability under certain loss functions and denoiser conditions.

The \textbf{motivations} of this work are: (1) Previous methods rely on regular grids or sequences, whereas NLOS requires flexible data representation due to its dependence on {\em geometrical structure}. Regular grids often lead to redundancy and overhead. (2) Objects can be seen as {\em sparse} parts; for instance, a marble statue's head, upper body, and limbs form a graph structure connected by joints. Graph neural network (GNN) methods can address this sparse structure. (3) The {\em coupled reconstruction} of albedo and depth information requires balanced optimization via hyper-parameters in a mixed loss function.
Specific contributions are as follows:
\begin{itemize}
\setlength{\itemsep}{-1pt}
\item Firstly, we introduce a dual-branch graph learning framework \xnet for NLOS reconstruction, incorporating a two-stage training mechanism, shown in Fig.~\ref{fig:types}(d), which decouples albedo and depth reconstruction to achieve the best results respectively.
\item Secondly, to effectively extract geometrical features, we develop a graph block and a channel fusion block specifically tailored for NLOS feature, where dense grid data is converted into a sparse graph structure of the objects.
\item Lastly, extensive experiments validate \xnet's robustness across various scenarios, achieving state-of-the-art performance including mainstream real-world data, with less GPU memory.
\end{itemize}

\section{Related Work}
\label{Sec:related}
\textbf{Deep NLOS Reconstruction:} Compared to traditional algorithms \cite{nam2021low,pei2021dynamic}, deep learning algorithms can learn scene priors, extract features, and reconstruct hidden objects. Hybrid methods \cite{meng_gap-net_2020,mou_deep_2022,wt_dense_2021,zhang_ista-net_2018} introduce several multi-scale iterative model-guided unfolding networks for Confocal NLOS (C-NLOS) reconstruction. Recent work \cite{su2023multi, ye2024plug} transforms traditional optimization into iterative learning but focuses solely on reconstruction performance without testing on corrupted cases. Inspired by but different from previous researches \cite{Neural_Transient2021,liu2021non,wt_dense_2021, mou_deep_2022, NeuS2023, zhu2023compressive}, our proposed \xnet develops a multi-scale graph-based network for C-NLOS reconstruction.
\begin{figure}[!t]
    \centering
 \includegraphics[width=0.9\linewidth]{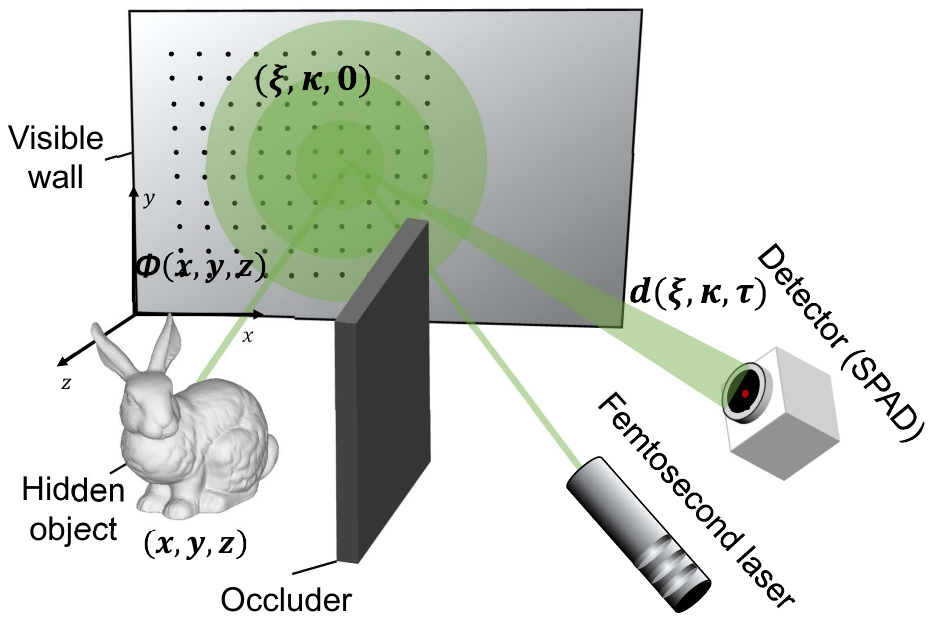}
    \caption{A schematic diagram of the C-NLOS system.}
    \label{fig:fig_forward}
\end{figure}

\begin{figure*}[htbp]
    \centering
    \subfloat[Framework]{\label{fig:frame}
    \hspace{-0.5em}
\includegraphics[width=0.7\linewidth,trim= 0 0 0 0,clip]{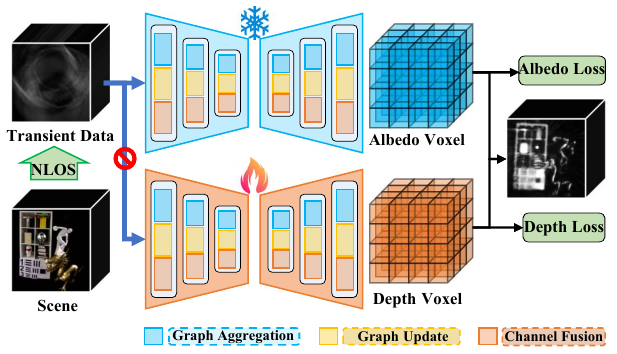}
    }
    \subfloat[Graph Module]{\label{fig:graph}
    \includegraphics[width=0.14\linewidth,trim= 0 0 0 0,clip]{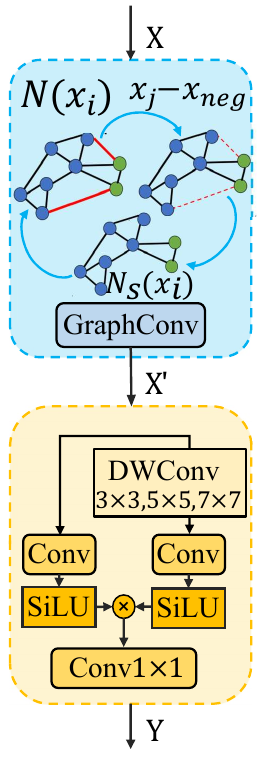}
    }
    \subfloat[Channel]{\label{fig:channel}
    \hspace{-0.5em}
\includegraphics[width=0.14\linewidth,trim= 1 8 0 1,clip]{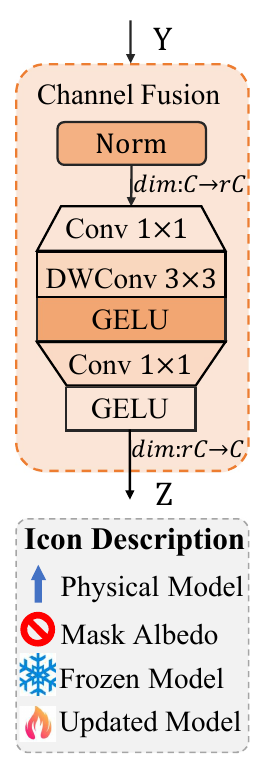}
    }
    \caption{
    \textbf{(a) Structure of the two-stage learning pipeline:}  The loss functions including Albedo and Depth are calculated in triple scales, respectively. Finally, the output voxels of two branches are combined for the optimized reconstruction. \textbf{(b) Structure of graph module.}
    \textbf{(c) Structure of channel fusion.} Each module adopts the Resnet skip connection mechanism.
    }
\end{figure*}

\noindent\textbf{Graph Learning:} GNNs are designed to process graph data by establishing long-range correlations in non-Euclidean space. Micheli~\cite{micheli2009neural} introduced the spatial graph convolutional network with nonrecursive layers. Instead of directly aggregating features from neighboring nodes, EdgeConv~\cite{wang2021object} obtains local neighborhood information by subtracting the central vertex's feature from that of neighboring vertices.  \cite{yang2020distilling} introduced a highway GNN for user geo-location in social media graphs, using `highway' gates to enhance gradient flow. Their research indicated a decline in performance beyond 6 layers. \cite{xu2017scene} proposed a \emph{Jump Knowledge Network} to determine graph neighbors for each node based on the graph's structure. GNN applications in computer vision~\cite{landrieu2018large} include point cloud classification, scene graph generation, and action recognition. Point clouds are 3D points typically collected by LiDAR scans. GCNs have been used for classifying and segmenting point clouds~\cite{landrieu2018large,edgeconv}.

\section{Physical Forward Model}
\label{Sec:model}
%

%
The C-NLOS imaging system typically consists of a scanning pulsed laser and a single photon time-resolved detector, which focuses on the same points on a diffuse reflective wall. As shown in Fig.~\ref{fig:fig_forward}, the directly illuminated points $(\xi, \kappa, 0)$ on the visible wall are considered as the sampling points, and the first diffuse reflection ray propagates to the points $(x,y,z)\in\Omega$ on the hidden object. The second reflected wave from the object, $\phiv(x,y,z)$, and the third reflections, which have the same reflective position $(\xi, \kappa, 0)$ as the first one, result in a time-resolved diffusive intensity $d(\xi,\kappa,t)$ received by the detector, where $t$ represents the time of photon flight between the first and third reflections. Finally, a three-dimensional (3D) light transient $\dv(\xi,\kappa,t)$ is measured by an $m \times m$ array sampling. As derived in \cite{otoole_confocal_2018}, the 3D continuous signal is formulated as
\begin{equation}
  \begin{split}
&\dv (\xi, \kappa, t) = \\
&\quad\textstyle \iiint _{\Omega}\frac{1}{\rv^4(x-\xi, y-\kappa, z)} \phiv(x, y, z) \delta(2 \rv-c t) d x d y d z, \\ 
&\rv(x-\xi, y-\kappa, z) =\sqrt{(x-\xi)^2+(y-\kappa)^2+z^2}, \label{forward}
\end{split}
\end{equation}
where the Dirac delta function $\delta$ models the light propagation, $c$ is the speed of light. 
Note that $\rv$ is the distance between the sampling points and the corresponding points on the surface of the hidden object. Combining all the detected photon arrival events into a single histogram results in a discrete inhomogeneous variable as
%
\begin{equation}
\yv=\Amat\xv + \bv, 
\label{eq:forward}
\end{equation}
where $\yv \in \mathbb{R}^{n_x n_y n_t}$ represents the discretized measurement by scanning point $(n_x, n_y)$ with respect to the discretized time bin $n_t$. $\Amat\in \mathbb{R}^{n_x n_y n_t \times n_x n_y n_z}$ is the discretized version of the volumetric Albedo model in~\eqref{forward}. $\xv \in \mathbb{R}^{n_x n_y n_z}$ represents the discretized Albedo of the hidden object. $\bv \in \mathbb{R}^{n_x n_y n_t}$ denotes the dark count of the detector and background noise \cite{7283534}.

\section{Proposed \xnet}
Graphing is an effective\cite{jia2022visual} technique for preserving structural features and propagation knowledge during occluded object reconstruction. A key component of our \xnet is the graph blocks, which generate learnable parameters and guide the model in the NLOS reconstruction process. Here, we outline the \xnet  pipeline:

\subsection{Network Framework}
Following learning-based methods~\cite{chen2020learned, mu_physics_2022, li_nlost_nodate}, we transform spatial-temporal transient data to the 3D spatial domain grid feature $\Fmat$ using a physics-based prior in the feature transform layer following \cite{chen2020learned}.
The proposed \xnet features a dual-branch~\cite{FSINet} symmetric architecture for efficient hierarchical representation learning, as shown in Figure~\ref{fig:frame}. Both the encoder and decoder networks have three scales, with the decoder progressively reconstructing hidden objects from the input grid features $\mathbf{F}$.

The upper albedo branch extracts albedo representation by disentangling features into texture and style components. We mask the albedo of grid feature $\Fmat$ by retaining the maximum value voxel along the $z$ axis to generate depth-focused grid feature $\Fmat'$. The lower depth branch is trained to suppress depth-irrelevant structures with the upper branch parameters frozen. Finally, the depth-focused voxel provides depth information and albedo-focused voxel provides albedo information to render the depth-focused voxel with the corresponding albedo value, followed by~\cite{ye2024plug} Algorithm 2.
Next, we describe graph block and core modules. 
\begin{figure}[!t]
    \centering
    \includegraphics[width=\linewidth]{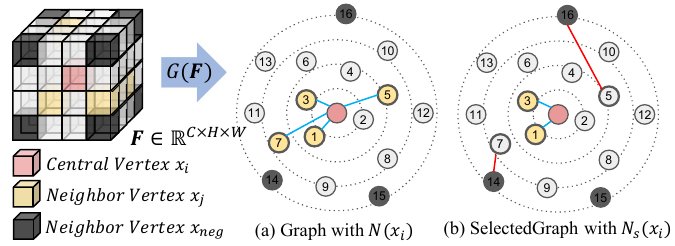}
    \caption{Graph construction and selection. \textcolor{cyan}{Cyan} edges connect neighbor vertices, while \textcolor{red}{red} edges denote negative connections. From (a) to (b), original neighbor vertices $7$ and $5$ are removed due to closer distance with negative vertices.}
    \label{fig:construct}
\end{figure}
\subsection{Graph Block}
The overall process of graph construction is defined as:
\begin{equation}
        \mathcal{G}=G(\Fmat) 
\end{equation}
Then, we explore the graph feature representations of NLOS. The graph aggregation block consists of
\begin{equation}
	\begin{split}
        &\mathcal{G}'= {F}(\mathcal{G}, \mathcal{W})\\
	&= {\rm Update}({\rm Aggregate}({\rm Select}(\mathcal{G}), W_{\rm agg}), W_{\rm update}),
	\end{split}
\end{equation}
where $W_{\rm agg}$ and $W_{\rm update}$ are the learnable weights of the aggregation and update operations, respectively. Concretely, the aggregation operation computes the representation of a vertex by aggregating features of neighbor vertices, and the update operation further merges the aggregated features.

\noindent\textbf{(1) Graph Construction} For a grid feature voxel $\Fmat$~$\in$~$\mathbb{R}^{H\times W \times C}$, we divided it into $N$ grids. 
By transforming each grid into a feature vector $\xv_i\in\mathbb{R}^{D}$, we have $\Xmat=[\xv_1,\xv_2,\cdots,\xv_N]$ where $D$ is the feature dimension and $i=1,2,\cdots,N$. These features can be viewed as a set of unordered vertices which are denoted as $\mathcal{V}=\{v_1,v_2,\cdots,v_N\}$. For each vertex $v_i$, we find its $K$ nearest neighbors $\mathcal{N}(v_i)$ and add an edge $e_{ji}$ directed from $v_j$ to $v_i$ for all $v_j\in\mathcal{N}(v_i)$. In this problem, we use $\xv$ instead of $v$ for convenience, and $\mathcal{N}(\mathbf{\xv}_i)$ is the set of NLOS voxel neighbor vertices of $\mathbf{\xv}_i$, shown in Fig.~\ref{fig:construct}(a). Here we adopt dynamic $k$-NN,~\cite{deepgcn} to construct the dynamic edges between points at every layer in the feature space. Then we obtain a graph $\mathcal{G}=(\mathcal{V},\mathcal{E})$ where $\mathcal{E}$ denotes all the edges. 

\noindent\textbf{(2) Graph Selection} To extract more effective information between vertices by aggregating features from its neighbor vertices, we define negative vertices $\xv_{neg}$ from the constructed graph and apply them to yield selected graph, shown in Fig.~\ref{fig:construct}(b):
\begin{align}
	\mathcal{N}_s(\xv_i) = \textstyle \max_k(\{\mathbf{\xv}_j-\mathbf{\xv}_{neg} | \xv_j \in\mathcal{N}(\mathbf{\xv}_i)\},
\end{align}
where $\xv_{neg}$ are the average pooling value of $8$ black vertices of the NLOS voxel, being the vertices with the least effective information. In our implementation, we compute a pairwise distance matrix in feature space and then take the $k$ vertices of maximum distance from $\xv_{neg}$ as the selected graph.

\noindent\textbf{(3) Graph Aggregation.}
Given the graph feature $\Xmat'\in\mathbb{R}^{D\times N}$ in Fig.~\ref{fig:graph}, which is resize to $\Xmat'\in\mathbb{R}^{D \times \sqrt{N} \times \sqrt{N}}$ for the aggregation operation:
\begin{equation}\label{eq:gcn}
	\Xmat'' = {\rm GraphConv}([\Xmat, \Xmat'] W_{in})W_{out} + \Xmat,
\end{equation}
where $\Xmat'\in\mathbb{R}^{kD \times N}$ is caculated by Eq.~(\ref{eq:cat}), $W_{in}$ and $W_{out}$ are the weights of fully-connected layers. After these layers, we resize the graph feature for the next operation.
\begin{align}\label{eq:cat}
	\mathbf{\xv}'_i = [\mathbf{\xv}_i,\{\mathbf{\xv}_j-\mathbf{\xv}_i| \xv_j \in\mathcal{N}_s(\mathbf{\xv}_i)\}],
\end{align}
and we test four types of GCN~\cite{deepgcn} \emph{ResEdgeConv}, \emph{GraphSAGE}, \emph{GIN} and \emph{Max-Relative} respectively.

\noindent\textbf{(4) Graph Update} We Further employ three different update operation with depth-wise convolutions (DWConv) layers in dilation ratios $d\in \{1,2,3\}$ in parallel to capture different weights respectively: Given the aggregated feature $\Xmat''\in \mathbb{R}^{D\times N}$, $\mathrm{DW}_{5\times 5, d=1}$ is first applied for low-order features; then, the output is factorized into ${\Xmat}_l \in \mathbb{R}^{D_l \times N}$, ${\Xmat}_m \in \mathbb{R}^{D_m \times N}$, and ${\Xmat}_h \in \mathbb{R}^{D_h \times N}$ along the channel dimension, where $D_l + D_m + D_h =D$; afterward, ${\Xmat}_m$ and ${\Xmat}_h$ are assigned to $\mathrm{DW}_{5\times 5, d=2}$ and $\mathrm{DW}_{7\times 7, d=3}$, respectively, while ${X}_l$ serves as identical mapping; finally, all the branches can be updated in parallel and concatenated as the final value: $\Xmat_{D} = \mathrm{Concat}(\Xmat_{l, 1:D_{l}}, \Xmat_{m}, \Xmat_{h})$. Multi-order update operation allows the model to update information in multiple representation subspaces for feature diversity.

Then, to alleviate over-smoothing phenomenon in deep GCNs~\cite{li2018deeper,oono2019graph}, we introduce more feature transformations and nonlinear activations in our block. We utilize the gating operation to adaptively fuse priors and structure features:
\begin{align}
    \label{eq:moga}
    \Ymat &= \mathrm{SiLU}\big( \mathrm{Conv}_{1\times 1}(\Xmat') \big) \otimes \mathrm{SiLU}\big( \mathrm{Conv}_{1\times 1}(\mathrm{DW}(\Xmat_D)) \big). \nonumber
\end{align}
The nonlinear activation function $\mathrm{SiLU}$ is inserted after graph aggregation to avoid layer collapse.

\subsection{Channel Fusion Module}
Prevalent architectures perform channel fusion mainly by two linear projections, \textit{e.g.,} 2-layer channel-wise MLP~\cite{iclr2021vit, nips2021MLPMixer} with a expand ratio $r$ or the MLP with a $3\times 3$ DWConv between~\cite{ aaai2022LIT, nips2022hilo}.
Due to information redundancy across channels~\cite{iccv2019GCNet, efficientnet, cvpr2020Orthogonal}, vanilla MLP requires a number of parameters ($r$ default to 4 or 8) to achieve expected performance.
To address this issue, most current methods directly insert a channel enhancement module, \textit{e.g.,} SE module~\cite{hu2018squeeze}, into MLP. Unlike these designs that require an additional MLP bottleneck, we introduce a lightweight channel fusion module $\mathrm{CF}(\cdot)$ to adaptively reallocate channel-wise features in high-dimensional hidden spaces.
As shown in Fig.~\ref{fig:channel}:

\begin{equation}
\begin{aligned}
    \Zmat' &= \mathrm{GELU}\left(\mathrm{DW_{3\times 3}}(\mathrm{Conv_{1\times 1}}(\mathrm{Norm}(\Ymat)))\right),\\
    \Zmat &= \mathrm{GELU}(\mathrm{Conv_{1 \times 1}}(\Zmat' )) + \Ymat.
\end{aligned}
\end{equation}
Concretely, GELU is used to gather and reallocate channel-wise information with the complementary interactions.

 \begin{figure}[!t]
    \centering
    \includegraphics[width=\linewidth]{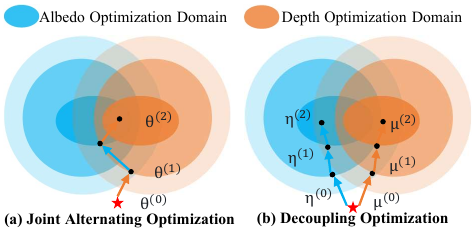}
    \caption{Illustration of joint optimization and decoupling optimization. $\thetav, \etav$ and $\muv$ are the optimized parameters for joint optimization, albedo, and depth branches, respectively.}
    \label{fig:dual}
\end{figure}

\subsection{Decoupled Optimization Strategy}
\label{sec:decouple}
It has been demonstrated in \cite{lu_depth_2020, depthnet_2021}  that structural information is closely associated with depth, whereas elements like style and lighting act as interference for depth perception.  To eliminate the impact of interfering information such as detailed textures, we develop a depth-focused branch to refine initial extracted features and focus on extracting graph structural representations to estimate depth in the second-stage training. 
\begin{align}
\etav^t \textstyle \leftarrow \argmin_\eta \mathcal{L}(\etav^{t-1}), \quad  \muv^t \textstyle \leftarrow \argmin_\mu \mathcal{L}(\muv^{t-1}), 
\end{align}
where $\etav, \muv$ are the optimized parameters for albedo and depth branches, respectively, shown in Fig.~\ref{fig:dual}(b). This design philosophy is inspired by~\cite{chen_exploring_2020}.  The alternating optimization provides an alternating trajectory, and the trajectory depends on the initialization. Starting from an arbitrary initialization, it may be difficult for the previous alternating optimization strategy to approach a joint optimal region for joint optimization (Fig.~\ref{fig:dual}(a)), but easier for our optimization strategy, shown in Fig.~\ref{fig:dual}(b).

\subsection{Multi-scale Loss Function}
\begin{figure}[!t]
    \centering
    \includegraphics[width=0.66\linewidth]{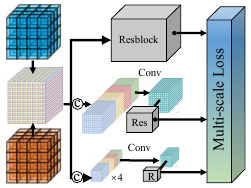}
    \caption{Illustration of our proposed multi-scale loss. Albedo voxel and depth voxel are sent to loss feedward respectively and trained in twstagesge.}
    \label{fig:loss}
\end{figure}

Taking into account the intricate albedo and depth information embedded within each voxel of NLOS, the {\em loss function} decouples the albedo loss and depth loss in the first train stage and second train stage, respectively.
As shown in Fig.~\ref{fig:loss}, a subset undergoes a stridden convolution, reducing it to a quarter of its original resolution. This architecture enhances performance by addressing multi-scale representation learning for various degradation, followed~\cite{MAXIM, cui2023selective}. 
In particular, this loss function encapsulates the diverse norm distances between the ground truth and the outputs across the triple-scale output:
\begin{align}
&\mathcal{L}_{\text {albedo }}=\textstyle \sum_{i=1}^3 \frac{1}{P_i}\|\mathrm{Albedo}(\hat{\mathbf{X}}_i)-\mathrm{Albedo}\left(\mathbf{X}_i\right)\|_1, \\
&\mathcal{L}_{\text {depth}}=\textstyle \sum_{i=1}^3 \frac{1}{P_i}\| \mathrm{Depth}(\hat{\mathbf{X}}_i)-\mathrm{Depth}(\mathbf{X}_i)\|_1, 
\end{align}
where $\mathrm{Albedo}({\Xmat_i})$ and $\mathrm{Depth}({\Xmat_i})$ is the 2D ground truth of albedo and depth value projected from 3D voxel. 
$\mathrm{Albedo}(\hat{\Xmat_i})$ and $\mathrm{Depth}(\hat{\Xmat_i})$ are projected from the output of two branches; Here we take the maximums for each pixel of the 3D voxel as the 2D albedo value, and their indexes multiplied by a corresponding distance coefficient as the 2D depth value following~\cite{chen2020learned}; $i$ is the index of multiple outputs, as illustrated in Figure~\ref{fig:loss}. $P_i$ represents the total elements of the output for normalization.

\section{Experimental Results}
This section details the implementation of our algorithm, presents simulation and real data results to show \xnet's superiority, and includes an ablation study for module evaluation. 
\subsection{Implementation Details}
\noindent\textbf{Methods for comparison.}
    PnP~\cite{ye2024plug},
    NLOST~\cite{li_nlost_nodate},
    MIMU~\cite{su2023multi} and 
    LFE~\cite{chen2020learned} are learning-based methods; Phasor~\cite{liu2019non}, LCT~\cite{otoole_confocal_2018},  
    SP~\cite{wu_nonline--sight_2021}, FK~\cite{lindell_wave-based_2019} and FBP~\cite{arellano_fast_2017} are non-learning-based methods.
    
\noindent\textbf{Training and testing datasets} For the practical C-NLOS reconstruction, our training dataset consists of 3000 generated measurements with corresponding intensity image and depth $256\times256$ pixel resolution 2D images, which involves various \texttt{bikes} 3D model. To ensure a fair comparison with existing methods, the training dataset, downloaded from a Google Drive link, is identical to that of LFE~\cite{chen2020learned}. The real data includes 6 scenes provided in \cite{lindell_wave-based_2019}.

\noindent\textbf{Training Strategy} The proposed method is implemented by the pytorch 1.7. The models are trained using Adam~\cite{kingma2014adam} with initial learning rate as $8e^{-4}$, which is gradually reduced to $1e^{-6}$ with cosine annealing~\cite{loshchilov2016sgdr}. Albedo branch is first trained on 8 samples for 150 epochs and depth branch is trained on 8 samples for 80 epochs. We utilize a Nvidia GeForce RTX 3090 to train and test the proposed model.

\subsection{\textbf Results on Simulated Datasets}
\begin{figure}[t!]
    \centering \includegraphics[width=\linewidth]{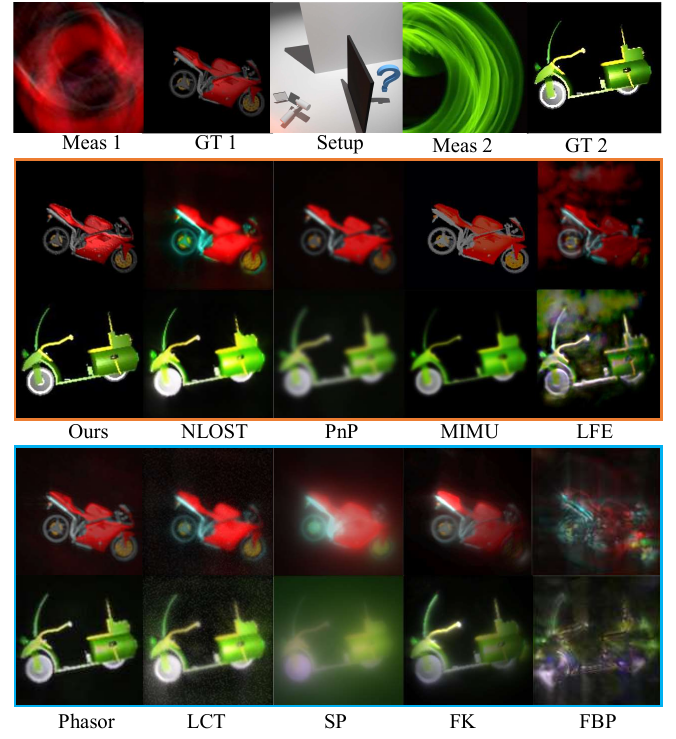}
    \caption{Qualitative visualization of two color samples in $256\times256\times512$ size reconstructed by the proposed \xnet and previous strong baseline methods. ``Meas'', and ``GT'' stands for measurement and ground truth respectively.}
    \label{fig:teaser}
\end{figure}

\noindent\textbf{Colorful Scale.} The proposed \xnet retrieves sharp shapes and colorful textures (Fig. \ref{fig:teaser}). All methods use the same physical parameters. SP uses 150 iterations. For LFE, FK performs better than LCT and Phasor. \xnet, NLOST, MIMU, LFE, and PnP achieve good geometric performance, though PnP distorts color. This shows \xnet's scalability and robustness across spectral channels. \xnet retains sharper edges and finer textures on motorbikes and avoids artifacts seen in other methods.

\begin{table}[t!]
    \centering
    \setlength{\tabcolsep}{0mm}
    \scriptsize
\begin{tabular}{l|ccccccc}
\toprule
Methods  & Venue&Type& PSNR(dB)$\uparrow$ & SSIM$\uparrow$ & RMSE$\downarrow$ & TIME(S)$\downarrow$  &Mem(G)$\downarrow$\\
\midrule
\textit{SP}  & Nat. Com.'2021&& 18.34 & 0.76 & 0.65 & -  &-\\
\textit{FBP} & Siggraph'2017&Non-& 17.63 & 0.62 & 0.73 & 0.65  &15.6\\
\textit{LCT} & Nature'2018&Learning- & 19.02 & 0.85 & 0.59 & 0.89  &17.7\\
\textit{FK} & TOG'2019&Based& 23.51 & 0.87 & 0.54 & 1.63  &21.0\\
\textit{Phasor} & Nature'2019& & 24.53 & 0.90 & 0.47 & 1.92  & 10.8 \\
\midrule
 \textit{LFE} & TOG'2020& & 27.54 & 0.89 & 0.08 &0.96  & \textbf{5.0} \\
 \textit{MIMU} & PG'2023& Learning-& 28.77& 0.92& 0.06&1.12 & 5.8 \\
 \textit{NLOST} & CVPR'2023 & Based & 28.17 & 0.90 & 0.07 & 0.24 & 20.3 \\
 \textit{PnP} & TOG'2024&  & 25.37& 0.83& 0.22&3.70 & 13.7 \\
 \rowcolor[rgb]{0.898,0.898,0.902}
 \textbf{DG-NLOS}&Ours& & \textbf{29.93}&  \textbf{0.92}& \textbf{0.04}&\textbf{0.18} & 9.6 \\
 \bottomrule
\end{tabular}%
\caption{Quantitative results on the motorbikes dataset.}
\label{tab:psnr}
\end{table}

\noindent\textbf{Gray Scale.} To compare the C-NLOS reconstruction results fairly in the same size with existing methods, we evaluate the performance in quality and quantity on the same test data. We quantitatively evaluate the results among these baseline approaches in Tab.~\ref{tab:psnr}, where we can see that the accuracy of the proposed \xnet exceeds existing methods.

\begin{figure}
    \centering
    \includegraphics[width=\linewidth]{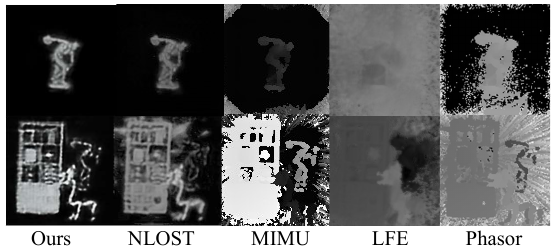}
    \caption{Depth estimation on real dataset. Brighter regions are closer to the observer than darker regions.}
    \label{fig:error}
\end{figure}

\begin{figure*}[!htbp]
    \centering    \includegraphics[width=\linewidth]{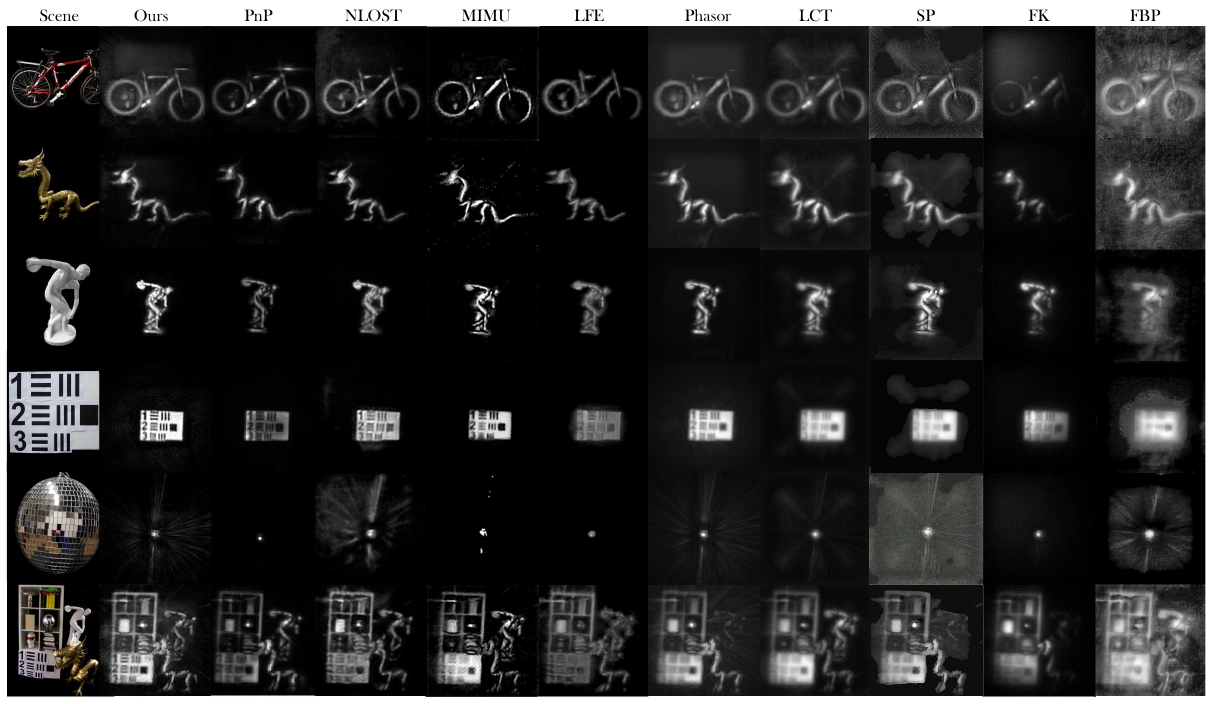}
    \caption{Reconstructions from real measurements at $32ps$ time resolution. Zoom in for more details.}
    \label{fig:real}
\end{figure*}
\noindent\textbf{RMSE Evaluation} Compared with the ground truth, the proposed method generates the least distance error and sharpest edges. The RMSE in Tab.~\ref{tab:psnr} shows that it especially outperforms existing methods thanks to the decoupled albedo-depth training strategy. 

\subsection{\textbf Results on Real Data}
\noindent\textbf{Albedo Reconstruction.} The results in Fig.~\ref{fig:real} using data from~\cite{otoole_confocal_2018} show that \xnet restores hidden details, high-frequency textures, and weak illumination, achieving superior quality. \xnet reconstructs object surfaces with sharper, cleaner details, including the bike front wheel, dragon tail, and bookshelf, surpassing previous methods. In the bottom row of Fig.~\ref{fig:real}, the statue is hard to reconstruct with existing methods due to its distance from foreground objects. Our graph structure dynamically adjusts geometric weights, enhancing object perception.

\noindent\textbf{Depth Estimation.} Most deep learning methods can achieve good results on simulated data sets, shown in Tab.~\ref{tab:psnr}, but perform poorly on the real data sets, shown in Fig.~\ref{fig:error}. We show the depth estimation from Phasor for comparison. LFE only estimates outlines of some objects; MIMU makes an error in estimating the front and rear distance of an object because the bookshelf should be between the closest dragon and furthest statue; NLOST estimates depth in details, but fails to eliminate the influence of texture with number on the board. Our \xnet can revise it with fewer artifacts due to the decoupled strategy analyzed before. 

\subsection{Ablation Study}
\label{sec:ablation}

\noindent\textbf{Effects of Graph and Channel Block.}
We first add the graph selection module and the graph update module to the baseline framework in Tab.~\ref{tab:ablation1}, which leads a 1.04dB increase in performance. We found that all proposed modules yield improvements with favorable costs.

\begin{table}[t]
    \setlength{\tabcolsep}{1mm}
    \centering
        \begin{tabular}{l|cccc}
        \toprule
        Modules & PSNR  &SSIM& Params. (M) & FLOPs (G) \\ \hline
        Baseline& 28.37 &0.86& 15.46& 146.06\\
        +$\mathrm{Select(\cdot)}$& 28.68 &0.89& 15.46& 146.92\\
        +$\mathrm{Update(\cdot)}$& 29.41 &0.90& 15.84& 147.16\\
        +$\mathrm{Fusion(\cdot)}$& \bf{29.93} &\bf{0.92}& 16.04 & 148.02 \\ \hline
        \end{tabular}
    \caption{Ablation of modules: Baseline uses \emph{ResEdgeConv}, MLP as $\mathrm{Update}(\cdot)$ and  $\mathrm{Fusion} (\cdot)$ following~\cite{deepgcn}.
    }
    \label{tab:ablation1}
\end{table}

\begin{table}[t]
\begin{center}
\setlength\tabcolsep{0.5mm}
\begin{tabular}{cccccc}
\toprule
 & Method & PSNR  &SSIM
& Params. (M) & FLOPs (G) \\ \midrule
(a) & \textit{GIN} & 27.37 &0.82& \bf{15.35}& \bf{140.58}\\
(b) & \textit{GraphSAGE} & 27.68 &0.87& 16.38& 151.30 \\
(c) & \textit{Max-Relative} & 28.42  &0.91& 16.00& 142.62 \\
\rowcolor[rgb]{0.898,0.898,0.902}
(d) & \textit{ResEdgeConv} & \textbf{29.93}  &\bf{0.92}& 16.04& 148.02 \\
\bottomrule
\end{tabular}
\end{center}
\caption{Comparison with various Graph Aggregation.}
\label{other-att}
\end{table}

\noindent\textbf{Comparison with alternative Graph Aggregation.} We further demonstrate the superiority of our Graph Aggregation by replacing it with three popular graph aggregation methods.
As represented in Tab.~\ref{other-att}, the \textit{GIN} and \textit{GraphSAGE} lead to performance degradation. Considering the balance of performance and overload, \textit{ResEdgeConv}~\cite{deepgcn} is used in our final version.

\noindent\textbf{Effects of dual-branch framework.} We test the single branch in the same architecture and the training sequence in Tab.~\ref{tab:dual} and found that training albedo branch first and then training the depth branch can achieve the best performance.

\begin{table}
\centering
\begin{tabular}{c|cc|cc}
\hline \multirow{2}{*}{ Strategy } & \multicolumn{2}{|c}{ Albedo } & \multicolumn{2}{c}{ Depth } \\
\cline { 2 - 5 } & PSNR & SSIM & RMSE & MAD \\
\hline Single Branch & 28.28 & 0.88 & 0.07 & 0.04 \\
Depth First & 29.65 & 0.90 & $\textbf{0.03}$ & 0.02 \\
\rowcolor[rgb]{0.898,0.898,0.902}
Albedo First & $\textbf{29.93}$ & $\textbf{0.92}$ & $0 . 0 4$ & $\textbf{0.02}$ \\
\hline
\end{tabular}
\caption{Various training strategy. ``Single Branch'' adopts albedo-depth mixed loss function in~\cite{chen2020learned}.}
\label{tab:dual}
\end{table}

\begin{table}[t]
    \centering

        \begin{tabular}{ccc|ccc}
        \toprule
        Scale&  PSNR  
&SSIM& Type& PSNR  &SSIM\\ \hline
        1&  28.41&0.89& L1 & \bf{29.93} &\bf{0.92}\\
        2&  28.85&0.90& MSE & 26.90&0.86\\
        3&  \bf{29.93} &\bf{0.92}& L1+MSE& 28.63&0.91\\ \hline
        \end{tabular}
    \caption{Various loss scale (left) and types (right).
    }
    \label{tab:loss}
\end{table}

\noindent\textbf{Multi-scale Loss Function.} Tab.~\ref{tab:loss} left part shows how the multi-scale loss function influences the performance. We can find that performance increases with the larger number of loss scales, demonstrating the effectiveness of the proposed multi-scale loss framework.
which takes advantage of the global content awareness and local content interaction in different scales to enhance the perception of the weak signal.  
Furthermore, we investigate the effectiveness of the ``L1" by replacing it with ``MSE" and ``L1+MSE'' in Tab.~\ref{tab:loss} right part. showing the effect of our designs.

\section{Conclusion}
To our knowledge, we are the first to introduce a graph-based architecture, \xnet, designed to transform previous 3D grid features to flexible graph features, ensuring the reconstruction process is sparser than previous work. This method not only efficiently extracts geometric information, but also greatly reduces overload. The visual results and quantitative evaluations demonstrate that our proposed \xnet consistently surpasses existing learning-based and nonlearning-based methods. In the future, we aim to broaden the model's applicability to encompass a wider range of corruptions, generating universal and robust models. 

\newpage
\section{Acknowledgements}
This work was supported by The National Key R$\&$D Program of China (grant number 2024YFF0505603, 2024YFF0505600), the National Natural Science Foundation of China (grant number 62271414), Zhejiang Outstanding Youth Science Foundation (grant number LR23F010001), Zhejiang “Pioneer" and “Leading Goose" R$\&$D Program(grant number 2024SDXHDX0006, 2024C03182), the Key Project of Westlake Institute for Optoelectronics (grant number 2023GD007), the 2023 International  Sci-tech Cooperation Projects of the “Innovation Yongjiang 2035” R$\&$D Program (grant number 2024Z126).

\appendix

\bibliography{aaai25}

\end{document}